\documentclass[lettersize,journal]{IEEEtran}
\usepackage{amsmath,amsfonts}
\usepackage[ruled]{algorithm2e}
\usepackage{array}
\usepackage[caption=false,font=normalsize,labelfont=sf,textfont=sf]{subfig}
\usepackage{textcomp}
\usepackage{stfloats}
\usepackage{url}
\usepackage{amssymb}
\usepackage{verbatim}
\usepackage{soul}
\usepackage{graphicx}
\usepackage{cuted}
\usepackage{bm} 
\usepackage{multirow}
\usepackage{color}

\usepackage{cite}
\begin{document}
\title{Affine Frequency Division Multiplexing: Extending OFDM for Scenario-Flexibility and Resilience}

\author{Haoran Yin, \textit{Graduate Student Member, IEEE},  Yanqun Tang, Ali Bemani, \textit{Member, IEEE}, \\Marios Kountouris, \textit{Fellow, IEEE},  Yu Zhou, Xingyao Zhang, Yuqing Liu, Gaojie Chen, \textit{Senior Member, IEEE}, Kai Yang, \textit{Member, IEEE}, Fan Liu, \textit{Senior Member, IEEE}, Christos Masouros,  \textit{Fellow, IEEE}, \\Shuangyang Li, \textit{Member, IEEE}, Giuseppe Caire,  \textit{Fellow, IEEE}, and Pei Xiao, \textit{Senior Member, IEEE}
	\thanks{
		 Haoran Yin, Yanqun Tang (corresponding author), Yu Zhou, Xingyao Zhang, and Yuqing Liu  are with the School of Electronics and Communication Engineering, Sun Yat-sen University; 
		 
		 Ali Bemani is with the Mathematical and Algorithmic Sciences Laboratory, Huawei France Research and Development;
		 
		 Marios Kountouris is with the Communication Systems Department, EURECOM, and with the Andalusian Research Institute in Data Science and Computational Intelligence (DaSCI), Department of Computer Science and Artificial Intelligence, University of Granada;
		 	 
		 Gaojie Chen is with the School of Flexible Electronics (SoFE), Sun Yat-sen University; 
		 
		 Kai Yang is with the School of Information and Electronics, Beijing Institute of Technology; 
		 
		 Fan Liu is with the National Mobile Communications Research Laboratory, Southeast University;
		 
		 Christos Masouros is with the Department of Electrical and Electronic Engineering, University College London;
		 
		 Shuangyang Li and Giuseppe Caire are with the Department of Electrical Engineering and Computer Science, Technical University of Berlin;
		 
		 Pei Xiao is with the 5G/6G Innovation Centre, University of Surrey.	 
		}  
}
\markboth{}%
{Shell \MakeLowercase{\textit{et al.}}: A Sample Article Using IEEEtran.cls for IEEE Journals}
\maketitle

\begin{abstract}
\textbf{Next-generation wireless networks are conceived to provide reliable and high-data-rate communication services for diverse scenarios, such as vehicle-to-vehicle, unmanned aerial vehicles, and satellite networks. The severe Doppler spreads in the underlying time-varying channels induce destructive inter-carrier interference (ICI) in the extensively adopted orthogonal frequency division multiplexing (OFDM) waveform, leading to severe performance degradation. This calls for a new air interface design that can accommodate the severe delay-Doppler spreads in highly dynamic channels while possessing sufficient flexibility to cater to various applications. This article provides a comprehensive overview of a promising chirp-based waveform named affine frequency division multiplexing (AFDM). It is featured with two tunable parameters and achieves optimal diversity order in doubly dispersive channels (DDC). We study the fundamental principle of AFDM, illustrating its intrinsic suitability for DDC. Based on that, several potential applications of AFDM are explored. Furthermore, the major challenges and the corresponding solutions of AFDM are presented, followed by several future research directions. Finally, we draw some instructive conclusions about AFDM, hoping to provide useful inspiration for its development.         
}
\end{abstract}

\section{Introduction}
\begin{figure}[tbp]
	\centering
	\includegraphics[width=0.430\textwidth,height=0.443\textwidth]{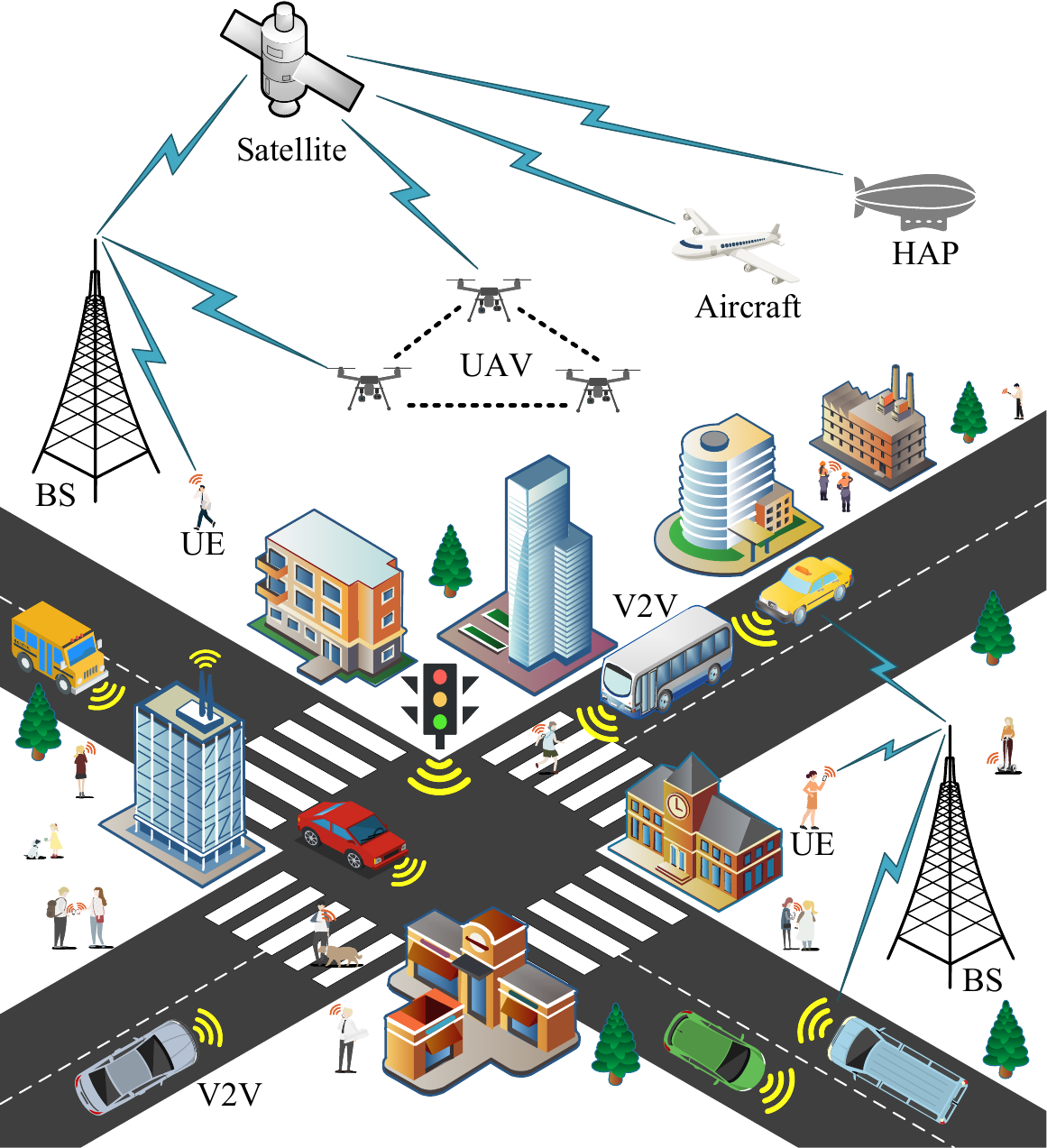}
	\caption{Visions of ubiquitous connection in NG.}
	\label{fig1}
\end{figure}

Next-generation (NG) wireless networks are envisaged to satisfy the pressing demands for ultra-reliable and wide-coverage communications in a variety of dynamic and high-mobility scenarios encompassing vehicle-to-vehicle (V2V), unmanned aerial vehicles (UAV), satellite networks, and so on, as illustrated in Fig. \ref{fig1}. The conventional fourth-generation (4G) and fifth-generation (5G) wireless networks, which consider the multipath effect in the approximated linear-time-invariant (LTI) channels as the major challenge, are no longer suitable for these highly dynamic scenarios. Specifically, the severe Doppler spreads caused by the relatively high mobility between the transmitter and receiver are expected to destroy the orthogonality among the subcarriers in orthogonal frequency division multiplexing (OFDM) waveform \cite{bb24.08.27.2}. Considering that higher frequency bands are likely to be used in the future to alleviate spectrum congestion, where the Doppler effect is more prominent, compensating the severe inter-carrier interference (ICI) in OFDM could not be a permanent solution. This encourages a new air interface design that can adapt to the heavy delay-Doppler (DD) spreads in doubly dispersive channels (DDC). Additionally, the flexibility and robustness of NG become increasingly indispensable to meet the various and diverse information interchange demands in the emerging era of ubiquitous intelligence. 

Against this background, many waveforms have sprung up in the past few years. One of the most influential schemes is a two-dimensional modulation waveform called orthogonal time frequency space (OTFS) modulation \cite{bb24.08.21.2}. By multiplexing the information symbols in the DD domain with the \emph{discrete Zak transform}, OTFS manages to capture the quasi-static nature of the DD domain representation of DDC. Another alternative is orthogonal chirp division multiplexing (OCDM), which is a chirp-based waveform with a fixed chirp rate determined by the \emph{discrete Fresnel transform (DFnT)} \cite{bb24.08.27.1}. While OCDM outperforms OFDM in many aspects, it is incapable of fully exploring the inherent diversity of DDC. Recently, a novel chirp-based multicarrier waveform termed affine frequency division multiplexing (AFDM) has been proposed for communications in high-mobility environments, showing great potential in realizing this goal and having advantages of lower channel estimation and multiuser multiplexing overhead over OTFS \cite{bb24.08.27.2}.

\begin{figure*}[htbp]
	\centering
	\includegraphics[width=0.9\textwidth,height=0.496\textwidth]{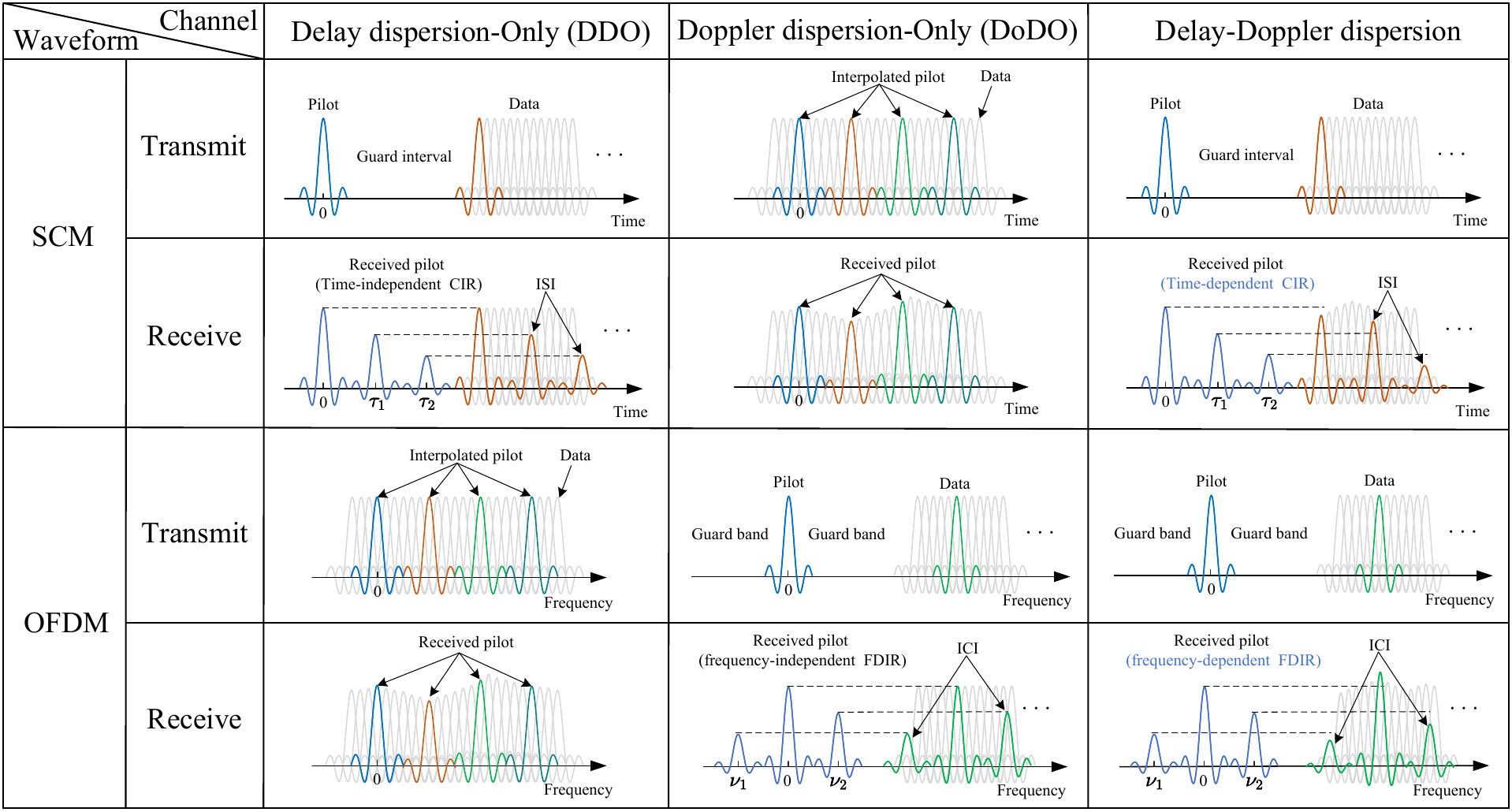}
	\caption{Influence of different dispersions on SCM and OFDM. The variables $\tau_{1}$ and $\tau_{2}$ denote two distinct delay shifts and the variables $\nu_{1}$ and $\nu_{2}$ denote two distinct Doppler shifts, satisfying $0<\tau_{1}<\tau_{2}$ and $\nu_{1}<0<\nu_{2}$, respectively.}
	\label{fig2}
\end{figure*}

The key idea of AFDM is modulating the information symbols on a set of orthogonal chirps via the \emph{discrete affine Fourier transform (DAFT)}, which is a generalization of the DFnT in OCDM and the \emph{discrete Fourier transform (DFT)} in OFDM \cite{bb24.9.23.3}. By tuning its two fundamental chirp parameters according to the DD profile of the DDC, AFDM is capable of separating the propagation paths with distinct delay or Doppler shifts in the underlying DAFT domain. This guarantees a quasi-static channel representation of DDC and enables low-overhead channel estimation. Moreover, the DD-path separability of AFDM also ensures its ability to harvest optimal diversity gain, which is important for ultra-reliable communications. Furthermore, AFDM’s flexibility, enabled by its two tunable parameters, enhances its suitability for diverse applications. These appealing advantages position AFDM as a strong contender for the NG waveform.

This article presents a comprehensive and accessible overview of AFDM. We begin with the challenges of traditional waveforms in DDC. Then, we study the fundamentals of AFDM, focusing on its tuneable chirp nature. The interactions between the chirp subcarriers and the DDC are graphically illustrated, reinforced by a discussion on the underlying principle of AFDM parameters selection. Additionally, several promising applications of AFDM are explored, highlighting its significant practical value. The main challenges associated with AFDM are then presented, along with effective solutions.  Finally, we outline several future research directions, aimed at sparking new ideas in this emerging field.

\section{Challenges in Doubly Dispersive Channels}
In general wireless channels, delay dispersion occurs when at least two taps exhibit different delay shifts, while Doppler dispersion occurs when at least two taps experience distinct Doppler shifts \cite{bb24.08.28.1}. Since the locations and velocities of various scatterers in real-world wireless channels are generally independent of each other, these two types of dispersion typically occur simultaneously.

Traditionally, single-carrier modulation (SCM) directly modulates information symbols in the time domain, as shown in Fig. \ref{fig2}. In delay dispersion-only (DDO) channels, the received signal is the superposition of $P$ delay-shifted replicas of the transmitted signal, where $P$ represents the number of paths. This leads to a convolutional input-output relationship (IOR), which inevitably introduces inter-symbol interference (ISI). Consequently, a time-domain pilot followed by a guard interval can be designed to estimate the channel impulse response (CIR), and the ISI can be further compensated by sequence-wise equalization. Moreover, in the case of Doppler dispersion-only (DoDO) channels, the $P$ Doppler-shifted replicas of the transmitted signal completely overlap at the receiver. This implies an element-multiplication (interference-free) IOR, allowing for the use of interpolated pilots for channel estimation, along with symbol-wise (single-tap) equalization. However, when it comes to doubly dispersive channels, the received signal undergoes both delay and Doppler shifts and hence the CIR is time-dependent. As a result, the CIR estimated from the pilot signal fails to correspond with that of the subsequent time-domain signals, hindering accurate equalization.

Correspondingly, traditional OFDM faces an analogous problem, as illustrated in Fig. \ref{fig2}. In DDO channels, the $P$ delay-shifted replicas of the transmitted signal entirely overlap at the receiver. Thus, an element-multiplication IOR is obtained, where interpolated frequency-domain pilots can be used to perform channel estimation, followed by low-complexity symbol-wise equalization. In contrast, in DoDO channels, the $P$ Doppler-shifted replicas of the transmitted signal spread across the frequency domain, causing serious ICI. Therefore, a frequency-domain pilot surrounded by guard bands can be adopted to estimate the Doppler shifts and channel gains of all the taps,  which constitute the frequency-domain impulse response (FDIR) and the ICI can subsequently be compensated with sequence-wise equalization. However, in the case of DDC, the FDIR is frequency-dependent, which means that the FIDR estimated by the pilot does not cohere with that of the signals residing in a different frequency. Therefore, effective equalization remains inaccessible in OFDM. In summary, both SCM and OFDM are incapable of managing the unavoidable interference in DDC, due to their limitations in interference estimation.

\section{Fundamentals of AFDM}
In this section, we introduce the basic concepts of AFDM, demonstrating its strong applicability over DDC.

\subsection{From OFDM to AFDM}
As outlined in the beginning, the surge in throughput demands from numerous emerging applications necessitates communication over high-mobility scenarios, where Doppler dispersion in conventional DDC becomes prominent. At the same time, low-dynamic scenarios will continue to play an important role, with OFDM remaining the preferred choice due to its low-complexity fast Fourier transform (FFT) aided modulation and frequency-domain single-tap equalization. Therefore, considering backward compatibility is crucial when evaluating waveform suitability, as it directly impacts the cost of upgrading the entire wireless network.

\begin{figure*}[htbp]
	\centering
	\includegraphics[width=0.995\textwidth,height=1.03\textwidth]{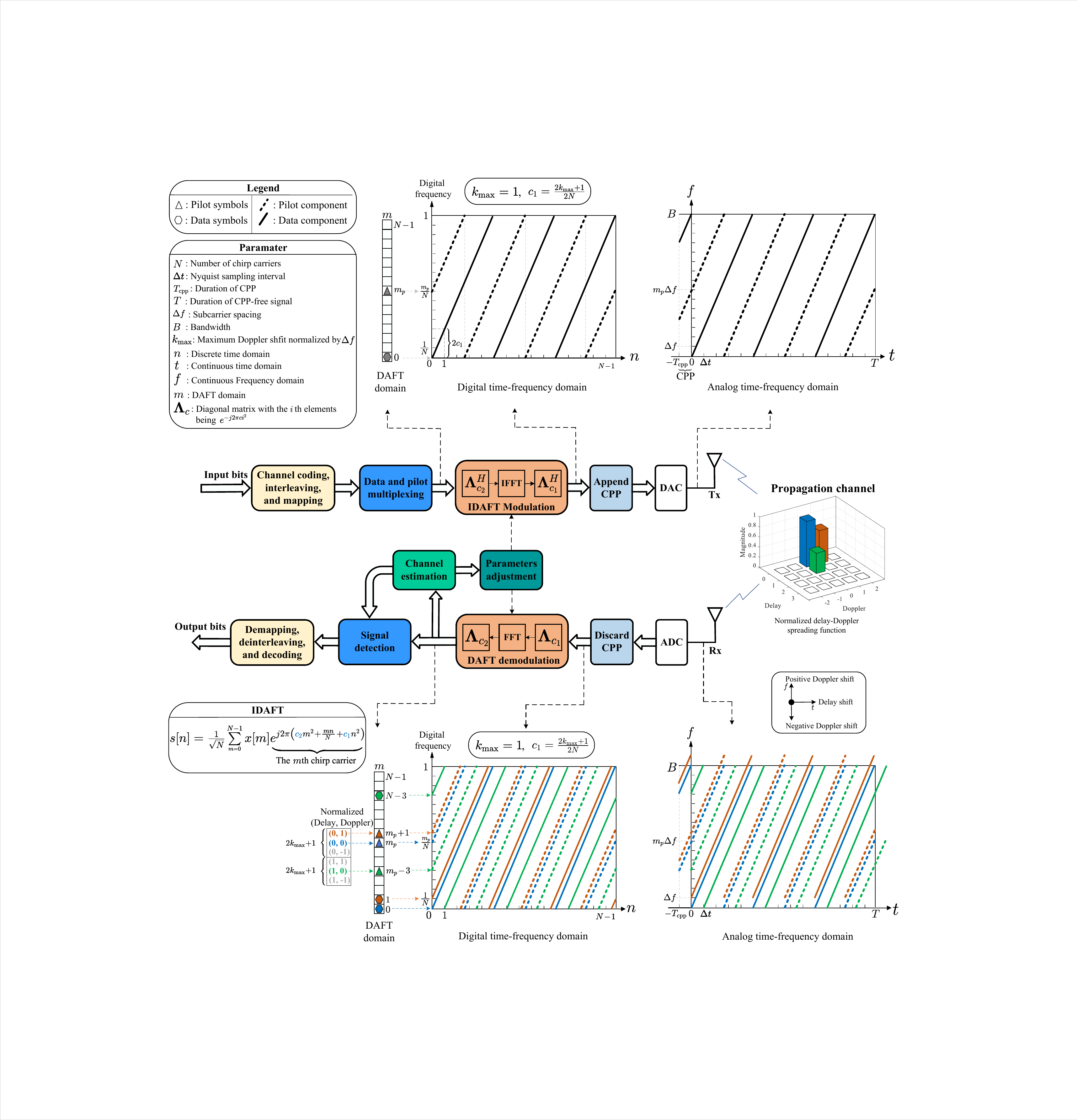}
	\caption{Baseband representation of AFDM transceiver architecture and its signal flow diagram.}
	\label{fig3}
\end{figure*}

\begin{figure*}[htbp]
	\centering
	\includegraphics[width=0.99\textwidth,height=0.303\textwidth]{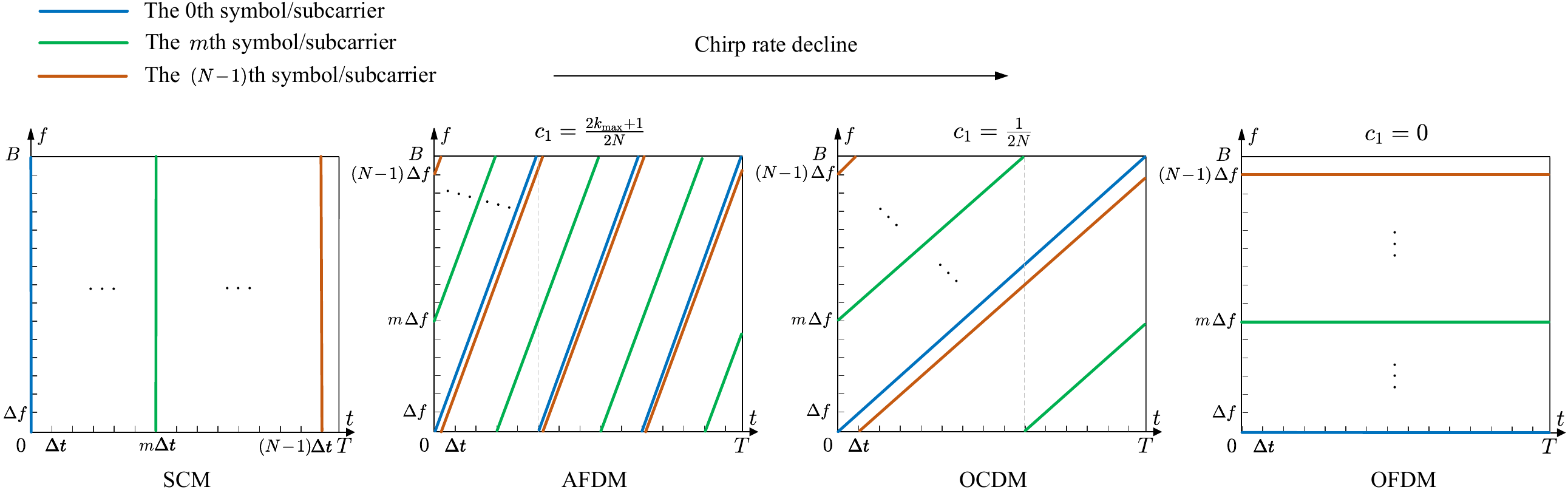}
	\caption{Time-frequency representations of SCM, AFDM, OCDM, and OFDM.}
	\label{fig4}
\end{figure*}

AFDM is a novel multicarrier waveform that employs well-designed chirp signals as subcarriers. By exploiting the inverse DAFT (IDAFT), whose exact expression is provided in Fig. \ref{fig3}, AFDM multiplexes $N$ information symbols onto $N$ orthogonal chirp subcarriers, respectively. Specifically, the variables $n$ and $m$ represent the discrete time domain and the chirp subcarrier index, respectively, with $m$ originally defined as the \emph{DAFT domain} in \cite{bb24.08.27.2}. Moreover, the two key parameters, $c_{1}$ and $c_{2}$, are highlighted in blue. In particular, $c_{1}$ is associated solely with the quadratic term of time $n^{2}$, which determines that all chirp subcarriers share the same digital chirp rate of $2c_{1}$. By contrast, $c_{2}$ is combined solely with the quadratic term of subcarrier index $m^{2}$, which implies that each chirp subcarrier has a different initial phase. Furthermore, the term $\frac{mn}{N}$ in IDAFT suggests a uniformly spaced initial frequency among all chirp subcarriers. These unique characteristics of AFDM are visualized in the time-frequency representation of its signal in Fig. \ref{fig4}. Therefore, as shown in Fig. \ref{fig3}, an $N$-length vector can be used to represent the information symbols in the DAFT domain, where each element therein corresponds to a chirp subcarrier. 

One appealing feature of the chirp subcarriers in AFDM is that they preserve their chirp characteristics even when affected by delay or Doppler shifts. This can be attributed to its time-frequency spanning nature, which provides it with the robust capability to perceive the delay shifts and Doppler shifts of the DDC simultaneously. Compared to the fully time-domain localized SCM symbol and the fully frequency-domain localized OFDM subcarrier shown in Fig. \ref{fig2}, the DD perceptibility of the chirp subcarriers is particularly attractive for communications over DDC. Based on that, the DD profile of the DDC can be effectively estimated in AFDM. For ease of understanding, a detailed AFDM transceiver diagram is provided in Fig. \ref{fig3}. At the transmitter, the data symbols are multiplexed in the DAFT domain, along with a pilot symbol. Here, to provide a straightforward insight into the interaction between the AFDM signal and the DDC, we only visualize the pilot symbol and one data symbol. The pilot symbol is positioned in the $m_{p}$th entry of the transmitted DAFT vector while the data symbol is arranged at the zeroth entry. After that, IDAFT is performed to obtain the discrete-time-domain AFDM signal, whose digital time-frequency-domain representation distinguishes the pilot and data components explicitly. Subsequently, a chirp-periodic prefix (CPP), due to the special chirp periodicity of AFDM chirp subcarriers, is appended to the start of the discrete-time-domain signal, serving the same purpose as the cyclic prefix (CP) in OFDM. Finally, after passing through a digital-to-analog converter (DAC), the analog AFDM waveform is obtained and is then transmitted into the DDC, which occupies a bandwidth of $B=\frac{1}{\Delta t}$ and a subcarrier spacing $\Delta f=\frac{B}{N}$, where $\Delta t$ represents Nyquist sampling interval. The total duration of the CPP-appended AFDM signal is $(T+T_{\text{cpp}})$, where the duration of the CPP-free part is $T=N\Delta t$ and the duration of the CPP part $T_{\text{cpp}}$ should be greater than or equal to the maximum delay shift of the DDC.

The received AFDM signal is composed of $P$ replicas of the transmitted AFDM signal. For clarity, an example of a three-path DDC is applied, with its normalized DD spreading function shown in Fig. \ref{fig3}, where the delay and Doppler shift are normalized by $\Delta t$ and $\Delta f$, respectively. In particular, the blue, brown, and green colors represent the paths with normalized (delay, Doppler) shifts of $(0,0)$, $(0,1)$, and $(1,0)$, respectively. Furthermore, as demonstrated in the analog time-frequency domain representation of the received signal, the three colored dotted chirp subcarriers conveying the pilot symbol clearly illustrate the impact of the DDC on the received AFDM signal. It is worth mentioning that the three colored solid chirp subcarriers bearing the data symbols establish the same relationship, given that all chirp subcarriers undergo the same wireless channel. After Nyquist sampling in the analog-to-digital converter (ADC) and discarding the CPP, we can acquire the received discrete time-domain AFDM signal. Next, DAFT-based demodulation is performed with the same $c_{1}$ and $c_{2}$ parameters as those in IDAFT to obtain the DAFT-domain receive symbols. We can notice that there are three pilot symbols, representing the three paths in the DDC. Therefore, the DDC can be estimated by first identifying the delay and Doppler shifts of all the paths according to the locations of the received pilot symbols. Afterward, the channel gains are calculated, based on which signal detection is performed.

In general, the DD perceptibility of chirp subcarriers in AFDM translates into the DD separability in the DAFT domain. It indicates that each DAFT-domain symbol in AFDM undergoes all propagation paths and can be sufficiently gathered at the receiver. Therefore, the signal-channel interaction is non-fading and the optimal diversity order can be achieved in AFDM by fine-tuning the parameters $c_{1}$ and $c_{2}$, where the optimal diversity order refers to the number of propagation paths that are separable in DD domain and is essential for achieving ultra-reliable communications. Moreover, the DAFT-domain separability in AFDM also implies a one-dimensional dispersion pattern, leading to a much lower channel estimation overhead compared to the inherent two-dimensional dispersion in the DD domain in OTFS. Furthermore, the IDAFT/DAFT operation can be implemented efficiently by concatenating a diagonal matrix at the front and back of the IFFT/FFT module in OFDM. Consequently, AFDM offers high backward compatibility with OFDM, making it a highly appealing solution for extensive future deployment at a low cost.

\subsection{Guideline for AFDM System Design}
The two inherent tunable parameters $c_{1}$ and $c_{2}$ grant AFDM a high degree of flexibility to accommodate diverse demands in various applications. In particular, the $c_{1}$ is of most importance as it determines the time-frequency distribution of the AFDM signal. When $c_{1}$ is set as the integer multiple of $\frac{1}{2N}$ and $N$ is even, CPP reduces to conventional CP in OFDM. Moreover, as illustrated in Fig. \ref{fig3}, a unit Doppler shift corresponds to a unit shift in the DAFT domain, while a unit delay shift corresponds to $2Nc_{1}$-step shift in the DAFT domain. Therefore, setting $|c_{1}|\geq \frac{2k_{\text{max}}+1}{2N}$ guarantees a bijective relationship between the DD domain and DAFT domain. Meanwhile, the parameter $c_{2}$ can be adaptively adjusted to further enhance the performance of AFDM systems. For example, when $c_{2}$ is chosen as either an arbitrary irrational number or a rational number sufficiently smaller than $\frac{1}{2N}$, the optimal diversity order can be ensured. By contrast, letting $c_{2}=0$ would lower the mod/demodulation complexity, strengthening its compatibility with OFDM \cite{bb24.9.08.2}. Besides, OCDM and OFDM can be regarded as special cases of AFDM with  $c_{1}$ and $c_{2}$ being set as specific values, as the time-frequency representations of SCM, AFDM, OCDM $(c_{1}=c_{2}=\frac{1}{2N})$, and OFDM $(c_{1}=c_{2}=0)$ shown in Fig. \ref{fig4}, where a trend of chirp rate decline can be clearly observed.

Apart from the two fundamental parameters, other system parameters, such as $T$, $B$, and $N$, should be selected properly in practice. Generally, the larger the $T$ is set, the smaller the subcarrier spacing $\Delta f = \frac{1}{T}$ becomes, which provides a better Doppler resolution in AFDM. However, considering the system latency requirement, $T$ should be confined to an acceptable value. Meanwhile, an even $N$ is preferred to convert CPP to CP, which facilitates the FFT operation and lowers the modulation complexity. Moreover, as shown in \cite{bb24.9.08.3}, the mod/demodulation complexity of AFDM can be further reduced when $N$ is chosen to satisfy several conditions.

\section{Prospective Application Scenarios}
In this section, we discuss several potential applications of AFDM in next-generation wireless networks.

\subsection{Space-Air-Ground Integrated Networks}
Space-air-ground integrated networks (SAGIN) have become an ambitious solution for global coverage and high-throughput communications. The multi-tier satellites in space networks, the high altitude platforms (HAP), aircrafts, and UAV in aerial networks, and the vehicle-to-everything (V2X) in densely deployed terrestrial networks enable an immersive experience within an intelligent society. All of these advanced applications rely on communication in wireless channels with large DD shifts, which can be effectively addressed by AFDM due to its robust DD resilience. Furthermore, the two key parameters of AFDM offer the flexibility needed to meet the diverse requirements across different parts of the multidimensional SAGIN, enabling a unified yet customizable air interface design for heterogeneous SAGIN.

\subsection{Underwater Acoustic Communications}
Underwater acoustic (UWA) communications serve as the cornerstone of the underwater Internet of Things (UIoT). General UWA channels are characterized by severe path loss, large delay spread, rapid time variations, and non-Gaussian noise. Moreover, the Doppler shift, which is proportional to the ratio of the relative transmitter-receiver velocity to the speed of sound, can be significant in UWA channels given that the speed of sound in water is significantly slower compared to the speed of electromagnetic waves in the air. Therefore, AFDM has great potential in UWA communications due to its strong ability to leverage DD diversity, which is beneficial in accommodating the challenging conditions of UWA channels. Meanwhile, the inherent DD sparsity of the UWA channels may enable relatively low-complexity signal detection in AFDM systems.

\subsection{High-Frequency-Band Communicatons}
The ever-growing throughput demands drive the exploration of higher frequency bands, such as millimeter wave frequency band and terahertz spectrum in communications, in which the Doppler effect will be more prominent. For example, for an OFDM system with a typical carrier frequency of 4 GHz and subcarrier spacing of 15 kHz, the Doppler shift corresponding to the relative speed of 135 km/h is 500 Hz, while the counterparts of 24 GHz and 77 GHz are around 3 kHz and 9.6 kHz, respectively. Meanwhile, the serious carrier frequency offset (CFO) and phase noise (PN) induced by the high-frequency oscillators also pose a great challenge to the OFDM system. By contrast, the inborn Doppler-resilient feature of AFDM can adapt to these hostile conditions. Moreover, the ability of AFDM to achieve optimal diversity grants it strong resilience against severe path loss.

\subsection{Security and Privacy Communications}
The process of information exchange is vulnerable due to the broadcast nature of wireless channels. As increasing amounts of data are transmitted over wireless networks, the security and privacy protection capabilities of the air interface become critically important. In this regard, AFDM stands out from its competitors by offering superior anti-interference and anti-eavesdropping capabilities. The chirp subcarrier adaptation in AFDM, first and foremost, ensures high resilience to various types of interference. Moreover, AFDM's two inherent key parameters offer unique flexibility to resist eavesdropping. Specifically, parameter $c_{1}$ is channel-dependent and will adjust adaptively according to the channel conditions between the legitimate transmitter and receiver. Moreover, parameter $c_{2}$ can act as a dynamic secret key to encrypt transmitted data, making it challenging for eavesdroppers to intercept the information.

\section{Challenges and Solutions}
Despite its promise, AFDM, as a burgeoning waveform, comes with its own set of challenges. In the following section, we identify four fundamental research problems in AFDM and suggest potential solutions to address them.

\begin{figure}[tbp]
	\centering
	\includegraphics[width=0.442\textwidth,height=1.205\textwidth]{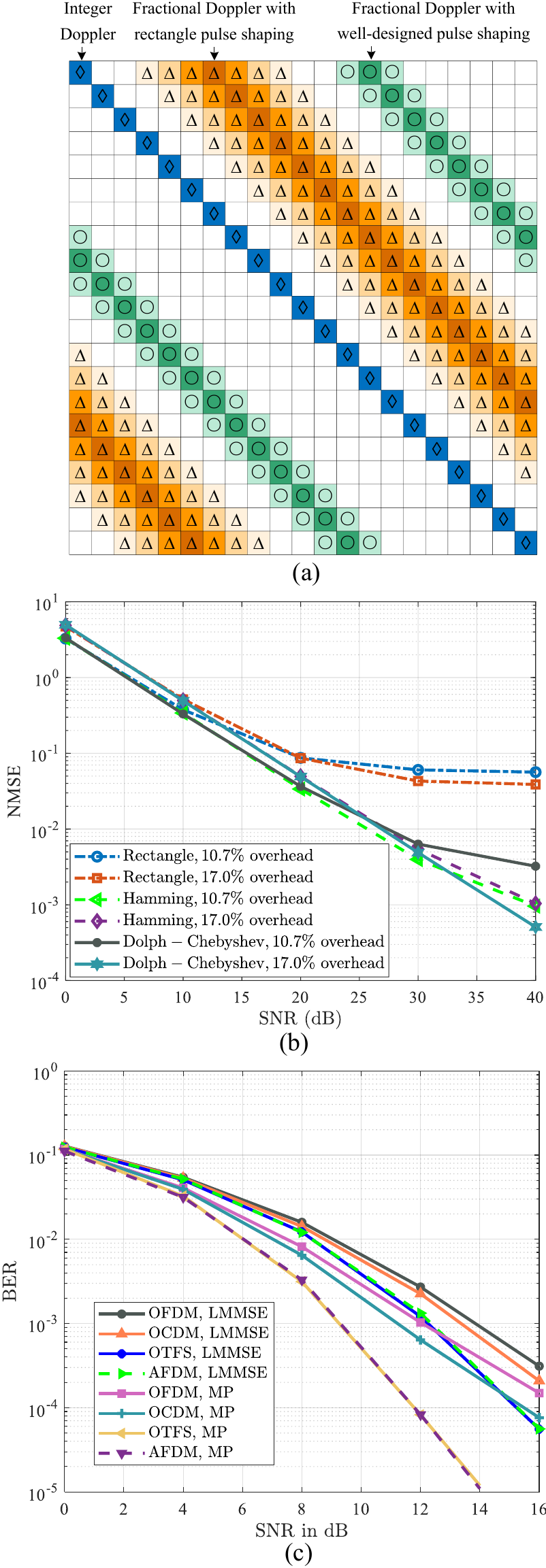}
	\caption{Performance of AFDM with different channel conditions, pulse shaping, and signal detectors:  a) Effective channel matrix of AFDM, three paths;
		b) Channel estimation NMSE versus SNR comparison with different pulse shaping, $N=512$, six paths with maximum delay of $3\Delta t$ and maximum Doppler of $2\Delta f$; c) BER versus SNR comparison among uncoded OFDM, OCDM, OTFS, and AFDM with different detectors, $N=512$, six paths with a maximum delay of $3\Delta t$ and a maximum Doppler of $2\Delta f$.}
	\label{fig5}
\end{figure}

\subsection{Channel Estimation}
Channel estimation is one of the most challenging tasks in practical communication systems, especially for high-mobility scenarios. Different from the conventional time or frequency domain channel estimation schemes, AFDM enables an embedded pilot-aided (EPA) channel estimation in the DAFT domain \cite{bb24.08.27.2}. The rationale behind this method is that the DAFT-domain channel representation of the DDC is DD separable, quasi-static, and compact, which can be interpreted by the bijective relationship between the DAFT domain and the DD domain channel representations. In particular, the DD separable and quasi-static properties ensure the channel estimability in the DAFT domain via a DAFT-domain pilot, as demonstrated in Fig. \ref{fig3}. Moreover, the compact characteristic of the DDC indicates a relatively small dispersion range in the DAFT domain, enabling the embedding of guard-protected pilots within data symbols, thereby enhancing spectral efficiency. Besides, it was shown in \cite{bb24.9.14.1} that the channel estimation overhead of AFDM can be lowered by carefully tuning the parameter $c_{1}$ when the DDC exhibits a certain degree of DD sparsity. 
 
While the aforementioned methods focus on how to acquire the delay shifts, Doppler shifts, and channel gains of all propagation paths in DDC, estimation errors are inevitable due to insufficient delay and Doppler resolutions of the AFDM signal. An effective way to tackle this issue is estimating the effective channel matrix (ECM) directly.  Fig. \ref{fig5}(a) illustrates the quasi-diagonal feature of AFDM ECM, which is determined by the underlying one-dimensional dispersion pattern in AFDM.  Instead of estimating the three channel parameters and then calculating the ECM, the embedded pilot-aided diagonal reconstruction (EPA-DR) scheme proposed in \cite{23.10.18.1} reconstructs the ECM directly from the received pilot symbols by exploring the diagonal reconstructability of AFDM ECM.

\subsection{Pulse Shaping}
The symbol spreading effect induced by fractional delay and Doppler shifts not only introduces ISI among pilot symbols but also causes ISI between the pilot symbols and the data symbols, which critically undermines the pilot-data separability at the receiver. Meanwhile, the symbol spreading effect also compromises the sparsity of the ECM, as demonstrated in Fig. \ref{fig5}(a), where the “$\diamond$” and “$\bigtriangleup$” components represent the path with integer Doppler and the path with fractional Doppler, respectively. An effective solution to solve this problem is suppressing the symbol spreading effect by performing pulse shaping and matched-filtering in the DAFT domain \cite{bb24.9.15.1}. Compared to the conventional rectangle pulse shaping with a high sidelobe level, applying a carefully-designed pulse shaping window with a lower sidelobe level, for example, Hamming or Dolph-Chebyshev, can significantly alleviate the symbol spreading effect. The notation “$\circ $” in Fig. \ref{fig5}(a) illustrates the suppression effect of a well-designed pulse shaping against fractional Doppler. Fig. \ref{fig5}(b) shows the normalized mean square error (NMSE) of the estimated ECM with different pulse shaping, where a more accurate ECM can be obtained with fewer guard symbols when the Hamming and the Dolph-Chebyshev pulse shaping windows are used compared to the conventional rectangle pulse shaping.

\subsection{Signal Detection}
After acquiring the ECM via channel estimation, signal detection should be conducted to recover the transmitted data symbols. Due to the inherent symbol dispersion in the DAFT domain, sequence-wise detection is necessary in AFDM. It means that dedicated detectors are needed to strike a compelling balance between the ability to harvest diversity gain and computation complexity, which are associated with the overall bit error ratio (BER) and detection efficiency, respectively. In this regard, we can explore the inherent properties of AFDM ECM to facilitate the signal detection process. For example, the quasi-diagonal structure of AFDM ECM can be explored via matrix decomposition to facilitate the matrix inversion operation in many non-iterative detectors, including zero-forcing (ZF) and linear minimum mean squared error (LMMSE) detectors. Moreover, iterative detectors based on message passing (MP) algorithm and its variants can be exploited to enhance the BER performance. In this case, the sparsity of the ECM is of significant importance since it reflects the degree of coupling among the data symbols at the receiver, which eventually determines the detection complexity. Fig. \ref{fig5}(c) shows the BER versus signal-to-noise ratio (SNR) comparison in uncoded OFDM, OCDM, OTFS, and AFDM systems with ideal channel state information and different detectors. We can observe that AFDM delivers comparable BER performance to OTFS and outperforms OCDM and OFDM significantly, which can be explained by the inherent optimal diversity order in AFDM.

\begin{figure}[tbp]
	\centering
	\includegraphics[width=0.48\textwidth,height=0.435\textwidth]{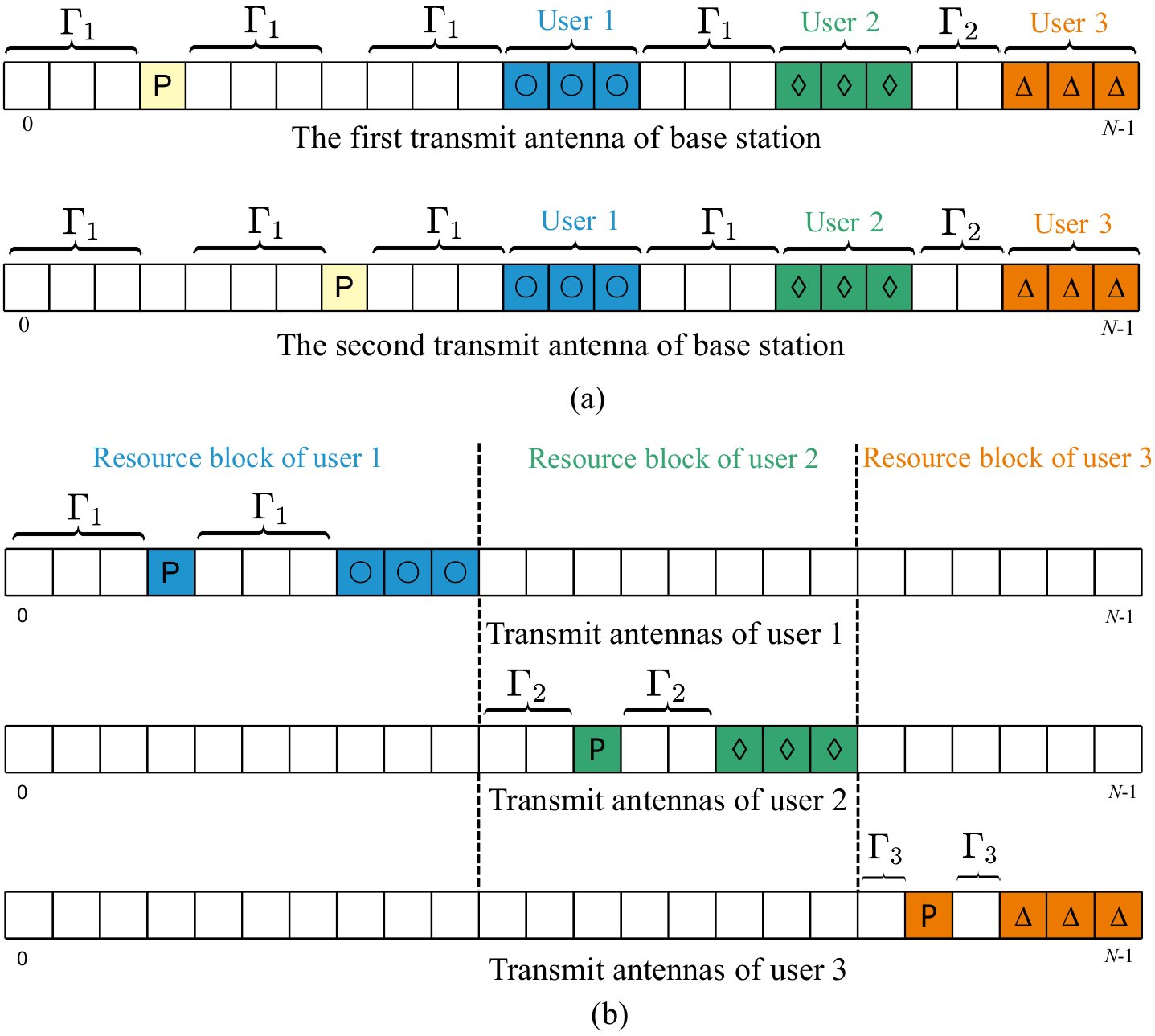}
	\caption{DAFT-domain resource allocation in AFDMA system (‘P’: pilot, ‘$\circ$’, ‘$\triangle$’, and ‘$\diamond$’: data symbols of the user 1,  user 2, and user 3, respectively): a) Downlink communications;
		b) Uplink communications.}
	\label{fig6}
\end{figure}

\subsection{MIMO and Multi-user Access}
Multiple-input multiple-output (MIMO) is a key enabler for ultra-reliable and high spectral efficiency communications. Therefore, a satisfying tradeoff between performance enhancement and complexity requirement should be carefully made to organically combine MIMO and AFDM. It was proven in \cite{23.10.18.1} that the optimal receive diversity order can be achieved by MIMO-AFDM, as well as linear increment in spectral efficiency with the number of transmit antennas when each transmit antenna transmits independent AFDM symbols. Moreover, a scalable space-time coding scheme called cyclic delay-Doppler shift (CDDS) can be applied to extract optimal transmit diversity gain in MIMO-AFDM. 

At the same time, a proper multiple-access scheme should be designed to support the massive connectivity requirements. Fig. \ref{fig6} illustrates an orthogonal resource allocation scheme in affine frequency division multiple access (AFDMA) system, where EPA channel estimation is also taken into account \cite{23.10.18.1}. Here, the differences in the DD profiles between the BS and all users are sufficiently explored by arranging the order of users according to their numbers of required guard symbols, which is denoted as $\Gamma_{i}$ for the $i$th user. This can significantly reduce the guard symbol overhead used to avoid inter-pilot-data interference and inter-user interference (IUI). Moreover, non-orthogonal multiple access (NOMA) can be applied to further enhance the spectral efficiency, for example, the sparse code multiple access (SCMA) studied in \cite{bb24.9.08.1}.

\section{Directions for Future Research}
In this section, we highlight several key open issues in AFDM that call for further investigations.

\subsection{Integrated Sensing and Communication}
Integrated sensing and communication (ISAC) is emerging as a key new feature of next-generation wireless networks. Chirp signals have been widely adopted in the area of sensing for a long time due to their strong ability to achieve long-range and high-resolution detection and robustness against various interference and noise. Therefore, the chirp-based AFDM waveform is naturally suitable for ISAC. While preliminary studies on AFDM-based ISAC design have established the great potential of AFDM in achieving sensing functionality without compromising its communications functionality \cite{bb24.03.15.5, bb24.9.08.4}, how to realize an optimal balance between the sensing and communication performances in a low-complexity manner remains a pressing issue that needs to be addressed.

\subsection{Full Duplex}
Full-duplex communication is a promising solution to alleviate the spectrum congestion problem in current wireless networks. It can theoretically double the spectral efficiency compared to the conventional time-division duplexing (TDD) and frequency-division 
duplexing (FDD) systems. To achieve this goal, stringent transmit-receive isolation is necessary. A prominent advantage of chirp-based waveforms is that full-duplex operation is potentially feasible if the received reflected signal arrives before subsequent chirps are transmitted. Therefore, how to efficiently implement full-duplex in AFDM systems over DDC is an exciting open area for further investigation.

\subsection{Coded-AFDM}
Channel coding is one of the most indispensable techniques for combating channel impairments and ensuring ultra-reliable communications. A fundamental tradeoff in AFDM shows that while the coding gain decreases at a diminishing rate as the number of separable paths increases, the diversity gain grows linearly. However, a dedicated modern channel code design for AFDM systems to sufficiently gather the optimal diversity gain with a practical detector and decoder is still missing in the literature. In this regard, joint detection and decoding along with a dedicated codebook design for coded-AFDM systems may be a promising solution to fully unleash the potential of AFDM.

\subsection{Numerology Design}
Numerology design is a key feature of the current 5G New Radio (NR), supporting diverse communication services that have very different requirements in terms of reliability, latency, and data rate \cite{bb24.9.25.1}. It has been standardized in the 3rd Generation Partnership Project (3GPP) and will continue to play a crucial role in NG, enabling flexible resource allocation. Therefore, how to design a scalable mix numerology family based on the CPP length, DAFT size, subcarrier spacing, and the two intrinsic chirp parameters of AFDM is an interesting topic to study. In particular, the low-complexity implementation and coexistence of multiple AFDM numerologies for various services should be taken into account carefully.

\subsection{Index Modulation}
Index modulation (IM) is a prospective method to improve the spectral and energy efficiency of conventional communication systems. Only a fraction of certain indexed resource entities, such as time slots, subcarriers, or antennas are activated in IM systems, while the remaining ones are left idle with their indices determined by extra information bits. Existing research on AFDM-IM has demonstrated its superiority over conventional OFDM-IM systems in DDC in terms of BER and spectral efficiency \cite{ bb24.9.23.2}. However, how to reliably and efficiently detect the information bits conveyed by the resource indices and conventional constellation symbols in AFDM-IM systems remains a key challenge for ongoing research. 

\section{Conclusion}
In conclusion, AFDM stands out as a promising candidate for next-generation wireless networks, thanks to its robust delay-Doppler resilience and the flexibility empowered by its two tunable chirp parameters. We started with an overview of the primary challenges in doubly dispersive channels, followed by the principles and system design of AFDM. Furthermore, the primary challenges of AFDM, including channel estimation, pulse shaping, data detection, MIMO, and multi-user access, were investigated, along with their corresponding solutions. Additionally, prospective application scenarios for AFDM were discussed, showing its versatility and substantial potential. Finally, several promising research avenues were outlined, with the aim of providing valuable insights and inspiration for its future development.

\vfill
\end{document}